\documentclass{article}
\usepackage{spconf,amsmath,amsfonts,amssymb,graphicx,bm,enumerate,pifont,adjustbox}
\usepackage[outline]{contour}
\usepackage[linesnumbered,ruled,vlined]{algorithm2e}
\usepackage{array}
\usepackage{multirow}

\makeatletter
\newcommand{\thickhline}{%
    \noalign {\ifnum 0=`}\fi \hrule height 1pt
    \futurelet \reserved@a \@xhline
}
\newcolumntype{"}{@{\hskip\tabcolsep\vrule width 1pt\hskip\tabcolsep}}
\makeatother

\makeatletter
\newcommand{\thealgorithm}{\arabic\algocf@float}

\newcommand{\AlgoCaptionFormat}{}

\renewcommand{\algocf@makecaption@ruled}[2]{%
  \global\sbox\algocf@capbox{\hskip\AlCapHSkip%
    \setlength{\hsize}{\columnwidth}
    \addtolength{\hsize}{-2\AlCapHSkip}
    \vtop{\AlgoCaptionFormat\algocf@captiontext{#1}{#2}}}
}%
\makeatother

\title{TRANSFORMER-BASED STREAMING ASR WITH CUMULATIVE ATTENTION}
\name{Mohan Li, Shucong Zhang, C\u{a}t\u{a}lin Zoril\u{a} and Rama Doddipatla}
\address{Cambridge Research Laboratory, Toshiba Europe Ltd, Cambridge, UK}
\begin{document}
\ninept
\maketitle
\begin{abstract}

In this paper, we propose an online attention mechanism, known as cumulative attention (CA), for streaming Transformer-based automatic speech recognition (ASR). Inspired by monotonic chunkwise attention (MoChA) and head-synchronous decoder-end adaptive computation steps (HS-DACS) algorithms, CA triggers the ASR outputs based on the acoustic information accumulated at each encoding timestep, where the decisions are made using a trainable device, referred to as halting selector. In CA, all the attention heads of the same decoder layer are synchronised to have a unified halting position. This feature effectively alleviates the problem caused by the distinct behaviour of individual heads, which may otherwise give rise to severe latency issues as encountered by MoChA. The ASR experiments conducted on AIShell-1 and Librispeech datasets demonstrate that the proposed CA-based Transformer system can achieve on par or better performance with significant reduction in latency during inference, when compared to other streaming Transformer systems in literature.

\end{abstract}
\begin{keywords}
End-to-end ASR, Transformer, online attention mechanism, cumulative attention
\end{keywords}
\section{Introduction}
\label{sec:intro}
In recent years, the ASR community has witnessed a surge in developing end-to-end (E2E) systems, which have exhibited less computational complexity and competitive performance when compared with the conventional hidden Markov model (HMM) based systems. For E2E ASR, the acoustic model (AM), lexicon and language model (LM) are integrated into a holistic network, and could be optimised without any prior knowledge. The E2E techniques can be generally categorised into three classes, known as connectionist temporal classification (CTC) \cite{graves2006connectionist, graves2014towards}, recurrent neural network transducer (RNN-T) \cite{graves2012sequence} and attention-based encoder-decoder framework \cite{chorowski2015attention, chan2016listen, chiu2018state}. 
As an epitome of the last class, Transformer \cite{vaswani2017attention} has been introduced to ASR following its overwhelming success in the field of natural language processing (NLP). Consequently, state-of-the-art performance is reported on a number of standard ASR tasks \cite{karita2019comparative}. The self-attention mechanism that underpins the Transformer architecture has soon replaced the dominant role of RNN in terms of both acoustic and language modelling, leading to the advent of Speech Transformer \cite{dong2018speech}, Transformer Transducer \cite{tian2019self,zhang2020transformer} and Transformer-XL \cite{dai2019transformer}. Unlike RNN structures that process the inputs in the temporal order, the self-attention module captures the correlation between the elements of any distance, which efficiently lifts the restriction imposed by long-term dependency.

Although Transformer has shown prominent advantage over the RNN based systems, it faces major difficulties in online ASR. Both the encoder and the decoder require access to the full speech utterance, leading to large latency and thus limiting the use for practical scenarios. In this paper, we focus on the streaming strategies in the decoder to address the latency in recognition. Previously, the following methods have been proposed and were shown to be effective for streaming Transformer architecture: (1) hard monotonic attention mechanisms, represented by the monotonic chunkwise attention (MoChA) \cite{chiu2017monotonic,tsunoo2019towards,inaguma2020enhancing} and its simplified variant, monotonic truncated attention (MTA) \cite{miao2020online,miao2020transformer}; (2) Motivated by the streaming characteristic of CTC technique, triggered attention (TA) \cite{moritz2020streaming} takes the spikes of CTC score as the boundaries for computing attention weights; (3) blockwise synchronous inference \cite{tsunoo2021streaming} performs standard attention independently on each input chunk, with the broken decoder states carried over to the succeeding chunks; (4) Continuous integrate-and-fire (CIF) \cite{dong2020cif}, along with the decoder-end adaptive computation steps (DACS) \cite{li2021transformer} and its head-synchronous version (HS-DACS) \cite{li2021head} interpret the online decoding as an accumulation process of attending confidence, which is halted when the accumulator exceeds a predefined threshold.

Among the aforementioned streaming methods, MoChA has been widely acknowledged, where it considers the ASR decoding as a monotonic attending process. The output is emitted around its acoustic endpoint. However, the output decision in MoChA is merely dependent on a single frame, adding possibilities for accidental triggering in complex speech conditions. Apart from this, in the context of multi-head Transformer, some of the heads might not capture valid attentions, and fail to truncate the decoding before the end of speech is reached. HS-DACS circumvents the problems of MoChA by: (i) transforming the decisions based on single-frame-activation into an accumulation of attending confidence; (ii) merging the halting probabilities to produce a same halting position for all attention heads in a decoder layer. Nonetheless, it is still not guaranteed that there is enough confidence to trigger the output, and an additional stopping criterion in the form of maximum look-ahead steps was introduced.

To overcome the above issues encountered by MoChA and HS-DACS, the paper proposes the cumulative attention (CA) algorithm. The CA-based attention heads collectively accumulate the acoustic information along each encoding timestep, and the ASR output is triggered once the accumulated information is deemed sufficient by a trainable device. The proposed algorithm not only involves multiple frames into the acoustic-aware decision making, but also achieves a unified halting positions for all the heads in the decoder layer at each decoding step.

The rest of the paper is organised as follows: Section 2 presents the architecture of Transformer ASR and some online attention approaches. Section 3 elaborates the workflow of the proposed CA algorithm. Experimental results are demonstrated in Section 4. Finally, conclusions are drawn in Section 5.

\section{Streaming Transformer ASR System}
\label{sec:format}

A typical Transformer-based ASR system consists of three components: front-end, encoder and decoder. The front-end is commonly implemented as convolutional neural networks (CNNs), which aims to enhance the acoustic feature extraction and conduct frame-rate reduction. The output of the CNN front-end enters into a stack of encoder layers, with each comprising of two sub-layers as a multi-head self-attention module and a pointwise feed-forward network (FFN). The decoder also stacks several layers and has a similar architecture as that of the encoder, except for an additional cross-attention sub-layer that interacts with the encoder states.

The attention modules in the Transformer architecture adopts the dot-product attention mechanism to model inter-speech and speech-to-text dependencies:
\begin{equation}
    \mathrm{Attention}(\mathbf{Q},\mathbf{K},\mathbf{V}) = \mathrm{softmax}(\frac{\mathbf{Q} \mathbf{K}^\mathrm{T}} {\sqrt{d_k}}) \mathbf{V}, 
    \label{eq: dotproductatt} 
\end{equation}
where $\mathbf{Q,K,V} \in \mathbb{R} ^{T/L\times d_k}$ denote the encoder/decoder states in the self-attention sub-layers, or $\mathbf{Q} \in \mathbb{R} ^{L\times d_k}$ denotes the decoder states, and $\mathbf{K,V} \in \mathbb{R} ^{T\times d_k}$ denote the encoder states in the cross-attention sub-layers, given $d_k$ as the attention dimension and $T$, $L$ as the length of the encoder and decoder states, respectively.

The power of the Transformer architecture also arises from the utilisation of multi-head attention at each sub-layer, where the heads project raw signals into various feature spaces, so as to capture the sequence dependencies from various aspects:
\begin{equation} \mathrm{MultiHead}(\mathbf{Q},\mathbf{K},\mathbf{V}) = \mathrm{Concat}(\mathrm{head}_1,...,\mathrm{head}_H) \mathbf{W}^O, \end{equation}
\begin{equation} \mathrm{where\;head}_h = \mathrm{Attention}(\mathbf{QW}^Q_h,\mathbf{KW}^K_h,\mathbf{VW}^V_h), \end{equation}
where $\mathbf{W}^{Q,K,V}_h \in \mathbb{R} ^{d_k \times d_k}$ and $\mathbf{W}^O \in \mathbb{R} ^{d_m \times d_m}$ represent the projection matrices, and $H$ is the number of attention heads given $d_m = H\times d_k$.

With regard to streaming ASR systems, the main challenge for Transformer is that it requires the full speech to perform the attention computations. At the encoder side, a relatively straightforward approach to overcome this issue is to splice the input features into overlapping chunks, and feed them to the system sequentially \cite{miao2020transformer}. However, such a strategy cannot be applied to the decoder side, because the attention boundaries of the ASR outputs might not be restricted within a single chunk. As a result, several online attention mechanisms have been proposed. Monotonic chunkwise attention (MoChA) was one of the first methods to address the problem. During decoding, the attending decision is made monotonically for each encoder state. Once a certain timestep is attended, a second-pass soft attention is conducted within a small chunk to serve as the layer outcome. Head-synchronous decoder-end adaptive computation steps (HS-DACS) was later proposed to take advantage of the history frames and better handle the distinct performance of the attention heads. The halting probabilities were produced and accumulated along the encoder states across all heads, and the output is triggered when the accumulation reaches a threshold. In the next section, we'll present the cumulative attention (CA) algorithm, which incorporates both the acoustic-aware decision of MoChA, and the head-synchronisation ability of HS-DACS.

\section{Proposed Cumulative Attention Algorithm}
\label{sec:ca}

\SetAlgoCaptionSeparator{.}
\begin{algorithm}[t]
\DontPrintSemicolon
\SetAlgoLined
\KwIn{encoder states $\bm{k}(\bm{v})$, decoder states $\bm{q}$, input length $T$, number of heads $H$.}
\textbf{Initialization:} $i=1$, $y_0=\langle sos \rangle$, $t_0=0$ \;
\While{$y_{i-1} \neq \langle eos \rangle$} {
    \For{$h = 1$ \textbf{to} $H$} {
        $\bm{c}^h_{i,0}= \bm{0}$ \;
    }
    \For{$j = 1$ \textbf{to} $T$} {
        \For{$h = 1$ \textbf{to} $H$} {
            $a^h_{i,j}=\mathrm{Sigmoid}(\frac{\bm{q}^h_{i-1} (\bm{k}^h_{j})^\mathrm{T}}{\sqrt{d_k}})$ \;
            $\bm{c}^h_{i,j} = \bm{c}^h_{i,j-1} + a^h_{i,j} \cdot \bm{v}^h_{j}$ \;
        }
        $\bm{c}_{i,j}=\mathrm{Concat}(\bm{c}^1_{i,j},...,\bm{c}^H_{i,j})$ \;
        $p_{i,j} = \mathrm{HaltSelect} (\bm{c}_{i,j})$ \;
        \If{$p_{i,j} \geqslant 0.5$} {
            break \;
        }
    }
    $\bm{c}_i := \bm{c}_{i,j}$ \;
    $t_i = \mathrm{max} (t_{i-1}, j)$ \;
    $y_i=\mathrm{OutputLayer}(\bm{c}_i)$ \;
    $i \mathrel{{+}{=}} 1$ \;
}
\caption{CA Inference for Transformer ASR}
\label{algo:inference}
\end{algorithm}

According to the investigation presented in \cite{inaguma2020enhancing}, the lower Transformer decoder layers tend to capture noisy and invalid attentions, which does harm to the general decoding especially for streaming scenarios. Similarly, we adopt the layer-drop strategy by applying the proposed CA algorithm just to the cross-attention module of the top decoder layer. Meanwhile, the rest of the layers are only equipped with the self-attention module and perform language modelling. The workflow of the algorithm for a certain decoding step $i$ is described as follows. 

Like all other online attention mechanisms, at head $h$, for each encoding timestep $j$, an attention energy $e^h_{i,j}$ is computed given the last decoder state $\bm{q}^h_{i-1}$ and the encoder state $\bm{k}^h_j$:
\begin{equation}
    e^h_{i,j} = \frac {\bm{q}^h_{i-1} (\bm{k}^h_{j})^T} {\sqrt{d_k}}.
    \label{eq:att_en}
\end{equation}
The energy is immediately fed to a sigmoid unit to produce a monotonic attention weight:
\begin{equation}
    a^h_{i,j} = \mathrm{Sigmoid}(e^h_{i,j}).
    \label{eq:att_w}
\end{equation}
The sigmoid unit is regarded as an effective alternative to the softmax function in the streaming case, which scales the energy to the range (0, 1) without accessing the entire input sequence. As opposed to MoChA and HS-DACS where the outcome of eq. (\ref{eq:att_w})
is directly interpreted as the attending/halting probability that dictates the output triggering, CA interprets it as the relevance of the encoder state to the current decoding step. This is the same as the standard attention mechanism in the offline system or the second-pass attention performed in MoChA.

Next, an interim context vector is generated at $j$ in an autoregressive manner:
\begin{equation}
    \bm{c}^h_{i,j} = \bm{c}^h_{i, j-1} + a^h_{i,j} \cdot \bm{v}^h_{i,j},
    \label{eq:int_cv}
\end{equation}
which carries all the processed acoustic information accumulated at the current timestep (the term $\bm{c}^h_{i, j-1}$ is discarded when $j=1$). Though HS-DACS performs accumulation at each decoding step in a similar way, it is with respect to the halting probabilities instead of acoustic information.

Inspired by HS-DACS, in order to force all the attention heads to halt at the same position, the interim context vectors produced by different heads are concatenated into a comprehensive one:
\begin{equation}
    \bm{c}_{i,j} = \mathrm{Concat} (\bm{c}^1_{i,j}, ..., \bm{c}^H_{i,j}).
    \label{eq:comp_cv}
\end{equation}

We introduce a trainable device, referred to as \emph{Halting Selector}, to determine whether to trigger the ASR output at each timestep, which is implemented as a single/multi-layer deep neural network (DNN) with the output dimension of one in our system. Then $\bm{c}_{i,j}$ is input to the DNN to calculate a halting probability:
\begin{equation}
    p_{i,j} = \mathrm{Sigmoid} (\mathrm{HaltSelect}(\bm{c}_{i,j}) + r +\epsilon),
    \label{eq:hp}
\end{equation}
which represents the likelihood of halting the $i^{th}$ decoding step at timestep $j$, provided the acoustic features accumulated so far by all the attention heads. The parameters $r$ denotes a bias term that is initialised to -4, and $\epsilon$ is an additive Gaussian noise applied only to training in order to encourage the discreetness of $p_{i,j}$.

Here, one may notice the major difference between CA and other streaming methods. In CA algorithm, the interim context vectors are computed first, based on which the ASR outputs will be activated. Whereas for MoChA and HS-DACS, the attending decision is taken first and then the final context vector is produced. If the decision is made inappropriately, then the corresponding context vector can contain invalid acoustic information for the decoder. 

Since the halting selector assigns hard decisions, making the system parameters non-differentiable, the training requires to take the expected value of $c_i$ by marginalising all possible halting positions. Hence, a distribution of the halting hypotheses is given as:
\begin{equation}
    \alpha_{i,j} = p_{i,j} \prod^{j-1}_{k=1} (1 - p_{i,k}),
    \label{eq:alpha}
\end{equation}
which is a simplified version of the counterpart calculated in MoChA \cite{chiu2017monotonic}. Finally, the expected context vector for the decoding step $i$ is computed as:
\begin{equation}
    \bm{c}_i = \sum^T_{j=1} \alpha_{i,j} \cdot \bm{c}_{i,j}.
    \label{eq:cv}
\end{equation}
It is important to note that HS-DACS doesn't calculate such an expectation in training, as the accumulation is truncated by a preset threshold and only the one-best context vector is derived.

During CA inference, $p_{i,j}$ is monotonically computed at each timestep from $j=1$, and the decoding step $i$ would be halted at the earliest $j$ where $p_{i,j} \geqslant 0.5$. The corresponding interim context vector $\bm{c}_{i,j}$ is selected and sent to the output layer to predict the ASR output. The pseudocodes of the above inference process are presented in algorithm 1.

One should be aware that although the CA algorithm adopts the Bernoulli sample process at single timesteps similar to what MoChA does, the halting decision is based on the whole encoding history. Unlike MoChA where individual heads detect separate halting positions, in CA all the attention heads simultaneously contribute to $\bm{c}_{i,j}$ and have a unified halting position. This makes it less vulnerable to the disturbance in single frames. Also, CA rules out the use of arbitrary accumulation threshold as does in HS-DACS, allowing the decoding to be halted flexibly in the weak attention scenarios.

\section{Experiments}
\label{sec:exp}

\subsection{Experimental setup}
\label{subsec:setup}
The proposed CA algorithm has been evaluated on the AIShell-1 Chinese task and the Librispeech English task. Following the standard recipes in ESPNet toolkit \cite{watanabe2018espnet}, speech perturbation is applied to AIShell-1, and SpecAugment \cite{park2019specaugment} is conducted on Librispeech. The acoustic features are 80-dimensional filterbank coefficients along with 3-dimensional pitch information. The vocabulary of the above datasets is 4231 Chinese characters and 5000 BPE tokenised word-pieces \cite{sennrich2015neural}, respectively.

Both tasks adopt a similar Transformer architecture. The front-end consists of 2 CNN layers, with each having 256 kernels with the width of $3\times3$ and a stride of $2\times2$ that reduces the frame rate by 2 folds. In order to deal with the online input, the 12-layer encoder takes the chunkwise streaming strategy as in \cite{miao2020transformer} where the sizes of the left, central and right chunks are \{64, 64, 32\} frames. At each encoder layer, the number of heads, attention dimension and the unit size of FFN are \{4, 256, 2048\} for AIShell-1, and \{8, 512, 2048\} for Librispeech. The decoder of the system stacks 6 layers with the same parameters as the encoder for each task, and there's only 1 DNN layer in the halting selector.

During training, the joint CTC/attention loss is utilised for multi-objective learning with the weight of $\lambda_{CTC}=0.3$. The learning rate (lr) of both tasks follows the Noam weight decay scheme \cite{vaswani2017attention} with the initial lr, warmup step and number of epochs set to \{1.0, 25000, 50\} for AIShell-1 and \{5.0, 25000, 120\} for Librispeech. As for inference is concerned, CTC joint decoding is carried out with $\lambda_{CTC}=0.3$ for AIShell-1 and $\lambda_{CTC}=0.4$ for Librispeech. An external LM trained with the texts of the training set is incorporated to rescore the beam search (beam width=10) hypotheses decoded by the system, where the LM is a 650-unit 2-layer Long Short Term Memory (LSTM) network and a 2048-unit 4-layer LSTM for AIShell-1 and Librispeech, respectively.

\subsection{Experimental results}
\label{subsec:results}

\begin{table}[t]
\centering
\caption{Character error rates (CERs \%) on AIShell-1.}
\begin{tabular}{llll}
\hline \thickhline
\multicolumn{1}{l}{Model}  &&  \multicolumn{1}{c}{dev} & \multicolumn{1}{c}{test} \\ \thickhline \hline
\multicolumn{3}{l}{Offline}                                                               \\ \hline
\multicolumn{1}{l}{Transformer \cite{karita2019comparative}}    & & \multicolumn{1}{c}{-}      & \multicolumn{1}{c}{6.7}       \\ \thickhline \hline
\multicolumn{3}{l}{Online}      \\ 
\hline
\multicolumn{1}{l}{MMA-MoChA Transformer \cite{inaguma2020enhancing}}      & & \multicolumn{1}{c}{-}      & \multicolumn{1}{c}{7.5}       \\
\multicolumn{1}{l}{BS-DEC Transformer \cite{tsunoo2021streaming}}    && \multicolumn{1}{c}{6.4}      & \multicolumn{1}{c}{7.3} \\
\multicolumn{1}{l}{MoChA Transformer}  && \multicolumn{1}{c}{6.4}  & \multicolumn{1}{c}{7.2}\\
\multicolumn{1}{l}{HS-DACS Transformer}  && \multicolumn{1}{c}{6.3}  & \multicolumn{1}{c}{7.0}\\
\multicolumn{1}{l}{\bf{CA Transformer}}  && \multicolumn{1}{c}{\bf{6.3}}  & \multicolumn{1}{c}{\bf{7.0}}\\ 
\hline \thickhline
\label{tab:aishell}
\end{tabular}


\centering
\caption{Word error rates (WERs \%) on Librispeech.}
\resizebox{\columnwidth}{!}{%
\begin{tabular}{lcccc}
\hline \thickhline
\multirow{2}{*}{Model}      & \multicolumn{2}{c}{dev}   & \multicolumn{2}{c}{test}     \\ \cline{2-5} 
                        & \multicolumn{1}{c}{clean}    & \multicolumn{1}{c}{other} & \multicolumn{1}{c}{clean}    & \multicolumn{1}{c}{other} \\ \thickhline \hline
\multicolumn{3}{l}{Offline}  \\ \hline
\multicolumn{1}{l}{Transformer (ours)}  & \multicolumn{1}{c}{2.4}  & \multicolumn{1}{c}{6.0} 
                                        & \multicolumn{1}{c}{2.6}  & \multicolumn{1}{c}{6.1}    \\ \thickhline \hline
\multicolumn{3}{l}{Online}   \\ \hline
\multicolumn{1}{l}{CIF \cite{dong2020cif}} & \multicolumn{1}{c}{-}   & \multicolumn{1}{c}{-} 
                                           & \multicolumn{1}{c}{3.3} & \multicolumn{1}{c}{9.6} \\
\multicolumn{1}{l}{Triggered attention \cite{moritz2020streaming}}    & \multicolumn{1}{c}{-}  & \multicolumn{1}{c}{-} 
                                        & \multicolumn{1}{c}{2.8}  & \multicolumn{1}{c}{7.2} \\
\multicolumn{1}{l}{BS-DEC Transformer \cite{tsunoo2021streaming}} & \multicolumn{1}{c}{2.5} & \multicolumn{1}{c}{6.8} 
                                                                  & \multicolumn{1}{c}{2.7}  & \multicolumn{1}{c}{7.1} \\
\multicolumn{1}{l}{MoChA Transformer}  & \multicolumn{1}{c}{2.8}  & \multicolumn{1}{c}{7.1}
                                       & \multicolumn{1}{c}{3.1}  & \multicolumn{1}{c}{7.4} \\
\multicolumn{1}{l}{HS-DACS Transformer}  & \multicolumn{1}{c}{2.4}  & \multicolumn{1}{c}{6.6}
                                         & \multicolumn{1}{c}{2.6}  & \multicolumn{1}{c}{6.6} \\
\multicolumn{1}{l}{\bf{CA Transformer}}  & \multicolumn{1}{c}{\bf{2.5}}  & \multicolumn{1}{c}{\bf{6.7}} 
                                         & \multicolumn{1}{c}{\bf{2.7}}  & \multicolumn{1}{c}{\bf{6.8}} \\
\hline \thickhline
\label{tab:librispeech}
\end{tabular}%
}
\vspace{-7mm}
\end{table}

Table \ref{tab:aishell} and \ref{tab:librispeech} demonstrate the ASR performance of the proposed system on AIShell-1 and Librispeech datasets in terms of character-error-rate (CER) and word-error-rate (WER) respectively. The reference systems are chosen to have similar Transformer architecture, input chunk sizes and external LM. Besides, for a fair comparison with CA, both the MoChA and HS-DACS based systems are trained with only one cross-attention layer ($D=1$), except for the HS-DACS on Librispeech which has three ($D=3$ as models with $D=1$ or 2 failed to converge well).

We observe that on both tasks, the CA system achieves better accuracy than the other systems in literature. With regard to the reproduced MoChA and HS-DACS models, on AIShell-1, CA obtains a relative gain of 2.8\% when compared with MoChA, and similar performance to HS-DACS. As for Librispeech is concerned, CA outperforms MoChA in both clean and noisy conditions with the relative gains of 16.1\% and 10.8\%, respectively. Moreover, CA still achieves comparable WERs to HS-DACS, given fewer cross-attention layers were used in the CA system.

\subsection{Latency measurement}
\label{subsec:latency}

The latency of inference has also been measured for the proposed algorithm together with MoChA and HS-DACS. Here we adopt the corpus-level latency as defined in \cite{inaguma2020minimum}, which is computed as the difference between the boundary of the right input chunk $\hat{b}^k_i$ where the halting position is located, and the actual boundary of the output token $b^k_i$ obtained from HMM forced alignment:
\begin{equation}
    \triangle_{corpus} = \frac{1}{\sum^N_{k=1} |\bm{y}^k|} \sum^N_{k=1} \sum^{|\bm{y}^k|}_{i=1} (\hat{b}^k_i - b^k_i),
    \label{eq:latency}
\end{equation}
where $N$ denotes the total number of utterances in the dataset, and $|\bm{y}^k|$ is the number of output tokens in each utterance. Since there might be ASR errors in the hypothesis sequence that will result in faulty latency computation, we only include $\hat{b}^k_i$ of the correctly decoded tokens in the above equation. Though this might lead to different denominators in eq. (\ref{eq:latency}), the comparison of latency is still reasonable given the similar ASR accuracy achieved by three systems. Also, as all the online attention mechanisms in our experiments are independently performed at each decoding step, it is not guaranteed that the halting positions are monotonic. Thus, when computing the latency, $\hat{b}^k_i$ is always synchronised to the furthest timestep ever seen in the decoding process (see Algorithm. 1 line 18).

\begin{table}[t]
\centering
\caption{Latency (frames) on AIShell-1 and Librispeech.}
\resizebox{\columnwidth}{!}{%
\begin{tabular}{lccccccc}
\hline \thickhline
\multirow{3}{*}{Model}  & \multicolumn{2}{c}{AIShell-1} &   & \multicolumn{4}{c}{Librispeech}  \\ \cline{2-3} \cline{5-8}
                        & \multirow{2}{*}{dev} & \multirow{2}{*}{test} &  & \multicolumn{2}{c}{dev} & \multicolumn{2}{c}{test}   \\ \cline{5-8} 
                        & & & & \multicolumn{1}{c}{clean}    & \multicolumn{1}{c}{other} & \multicolumn{1}{c}{clean}  & \multicolumn{1}{c}{other} \\ \thickhline
\multicolumn{1}{l}{offline}  & \multicolumn{1}{c}{232.8}  & \multicolumn{1}{c}{257.0} & & \multicolumn{1}{c}{497.9} 
                            & \multicolumn{1}{c}{440.7}
                           & \multicolumn{1}{c}{528.5}  & \multicolumn{1}{c}{463.8} \\ \hline
\multicolumn{1}{l}{MoChA}  & \multicolumn{1}{c}{90.1}  & \multicolumn{1}{c}{92.5} & & \multicolumn{1}{c}{295.2}  
                           & \multicolumn{1}{c}{256.4} & \multicolumn{1}{c}{303.1}  & \multicolumn{1}{c}{271.2} \\
\multicolumn{1}{l}{HS-DACS}  & \multicolumn{1}{c}{53.4}  & \multicolumn{1}{c}{54.2} & & \multicolumn{1}{c}{163.8}  
            & \multicolumn{1}{c}{145.0} & \multicolumn{1}{c}{156.1}  & \multicolumn{1}{c}{146.5} \\
\multicolumn{1}{l}{\bf{CA}}  & \multicolumn{1}{c}{\bf{52.8}}  & \multicolumn{1}{c}{\bf{51.8}} & & \multicolumn{1}{c}{\bf{68.1}}                             & \multicolumn{1}{c}{\bf{63.5}} & \multicolumn{1}{c}{\bf{68.5}}  & \multicolumn{1}{c}{\bf{65.5}} \\
\hline \thickhline
\label{tab:latency}
\end{tabular}%
}
\vspace{-4mm}
\end{table}

Table \ref{tab:latency} presents the latency levels of MoChA, HS-DACS and CA based systems evaluated on AIShell-1 and Librispeech datasets. In order to have a fair comparison to CA, in both MoChA and HS-DACS, the maximum look-ahead steps ($M$) is not applied during the decoding process. On AIShell-1, we observed that the latency levels of both systems seem to be reasonable when compared to the offline system, while on Librispeech, it was noticed that the latency levels were close to the offline system, due to the redundant heads that cannot capture valid attentions. In order to reduce the latency, a maximum look-ahead step of $M=16$ is imposed during recognition only for the Librispeech task. One can observe that CA (without $M$) achieves better latency levels than MoChA and HS-DACS on both AIShell-1 (without $M$) and on Librispeech (with $M$) .


The poor latency performance given by MoChA and HS-DACS might be explained by looking at the halting decisions of various heads to generate the ASR outputs, as shown in Fig. 1. One can observe from Fig. 1 (a), heads 2, 3 and 6 in MoChA are mostly unable to halt the decoding and have to rely on the truncation executed by the maximum look-ahead steps. Similarly, in Fig. 1 (b) the accumulation of halting probabilities in HS-DACS heads fails to exceed the joint-threshold (number of heads, 8) at certain decoding steps, making the inference process reach the end of speech at early stages. On the other hand, the halting decisions of CA are made monotonically and always in time as illustrated in Fig. 1 (c). Although CA might also have redundant heads, these heads can be backed up by the other functioning ones, since all of them are synchronised and the halting decision is based on the overall acoustic information.

\begin{figure}[t]

\begin{minipage}[t]{1.0\linewidth}
  \centering
  \centerline{\includegraphics[width=8.5cm]{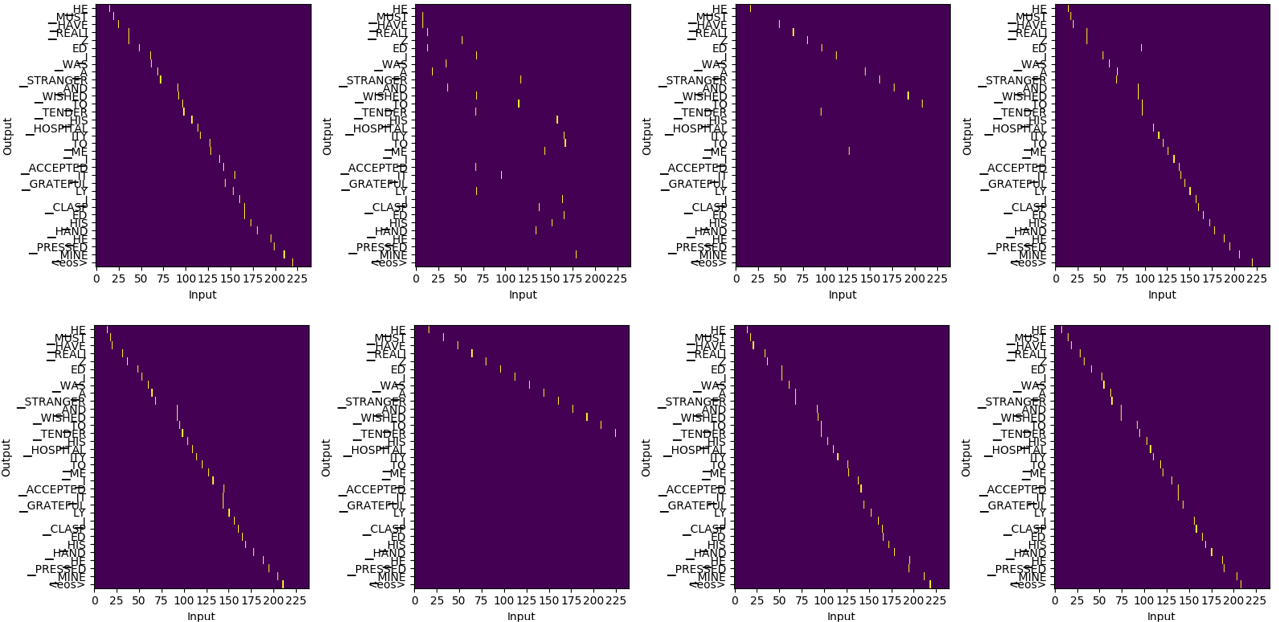}}
  \centerline{(a) MoChA}\medskip
\end{minipage}
\begin{minipage}[t]{0.6\linewidth}
  \centering
  \centerline{\includegraphics[width=3cm]{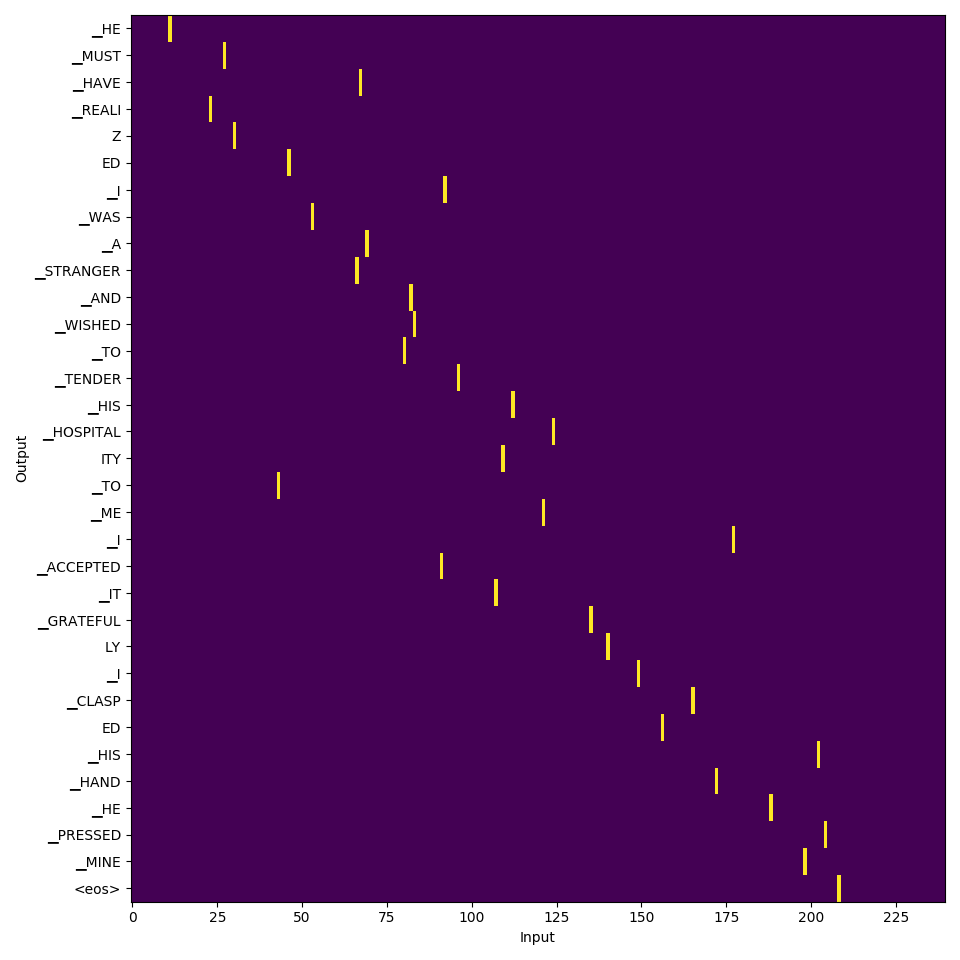}}
  \centerline{(b) HS-DACS}\medskip
\end{minipage}
\hskip -8ex
\begin{minipage}[t]{0.4\linewidth}
  \centering
  \centerline{\includegraphics[width=3cm]{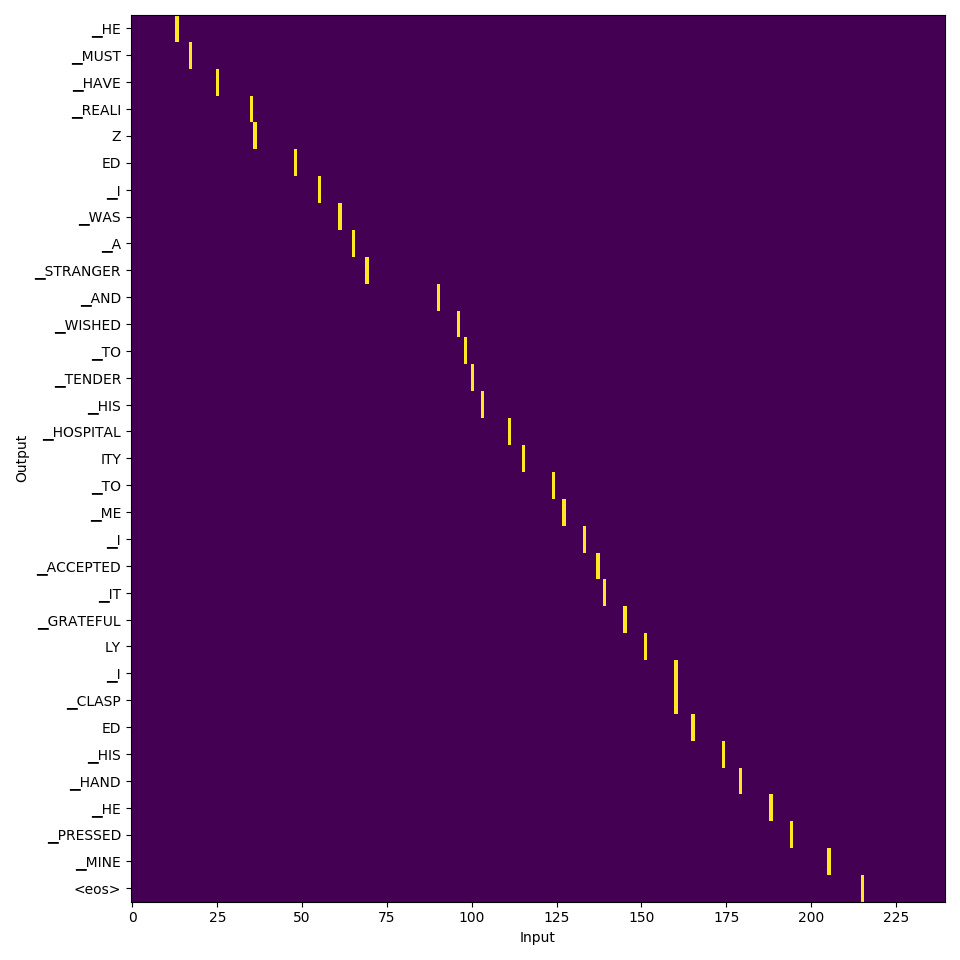}}
  \centerline{(c) CA}\medskip
\end{minipage}
\caption{Separate/joint halting decisions of the ASR outputs given by online attention heads in streaming Transformer systems.}
\label{fig:hp}
%
\end{figure}

\section{Conclusion}
\label{sec:conclu}

The paper presented a novel online attention mechanism, known as cumulative attention (CA) for streaming Transformer ASR. Combining the advantages of the acoustic-aware method (MoChA) and the accumulation based approach (HS-DACS), the CA algorithm utilised an trainable device called halting selector to determine robust halting positions to trigger the ASR output. The attention heads in the CA layer were synchronised to produce a unified halting positing in the decoder layer. By doing so, all the heads simultaneously contribute to the potential ASR output, so that the issues caused by the distinct behaviours of the heads were effectively eased off. Experiments on AIShell-1 and Librispeech showed that the proposed CA approach achieved similar or better ASR performance compared to existing algorithms in literature with significantly noticeable gains in latency during inference.



\bibliographystyle{IEEE}
\bibliography{refs}

\end{document}